\documentclass[10pt,twocolumn,cleanfoot,cleanhead]{config/asme2e}

\usepackage{empheq, amssymb}

\usepackage{lmodern}
\usepackage{amsmath}
\usepackage{amsfonts}
\usepackage{amssymb}

\usepackage{bm} 

\usepackage{txfonts}

\usepackage[T1]{fontenc}
\usepackage[utf8]{inputenc} 

\usepackage[dvipsnames]{xcolor} 

\usepackage[american]{babel} 

\usepackage{graphicx}

\usepackage{epstopdf}

\usepackage[noadjust]{cite} 

\usepackage{ltxcmds}

\usepackage{mathtools}

\usepackage{subcaption}

\usepackage{multirow} 

\usepackage{blindtext}

\usepackage{hyperref}

\usepackage{enumitem}

\usepackage{ifthen}

\usepackage{etoolbox}

\usepackage{cancel}

\usepackage{empheq}

\usepackage{fontawesome}

\usepackage{soul}

\usepackage{flushend} 

\usepackage{textcomp}

\usepackage{lipsum}

\definecolor{light-gray}{gray}{0.4}
\definecolor{box-gray}{gray}{1}

\hypersetup{
    unicode=false,          
    pdftoolbar=true,        
    pdfmenubar=true,        
    pdffitwindow=false,     
    pdfstartview={FitV},    
    pdftitle={On the Use of Geometric Deep Learning for the Iterative Classification and Down-Selection of Analog Electric Circuits},    
    pdfauthor={Anthony Sirico Jr. and Daniel R. Herber},     
    pdfsubject={},   
    pdfkeywords = {}, 
    pdfnewwindow=true,      
    colorlinks=true,
    allcolors=blue
}

\usepackage{nomencl}

\renewcommand\nomgroup[1]{%
  \item[\bfseries
  \ifstrequal{#1}{V}{ Variables}{%
  \ifstrequal{#1}{B}{ Subscripts}{%
  \ifstrequal{#1}{P}{ Notation}{%
  \ifstrequal{#1}{A}{ Acronyms}{}}}}]
}

\makenomenclature



\usepackage{dblfloatfix}
\usepackage{float}

\definecolor{block-gray}{gray}{0.95}

\usepackage[framemethod=TikZ]{mdframed}

\usepackage{xpatch}

\makeatletter
\xpatchcmd{\endmdframed}
  {\aftergroup\endmdf@trivlist\color@endgroup}
  {\endmdf@trivlist\color@endgroup\@doendpe}
  {}{}
\makeatother

\newcommand{\eref}[1]{Eq.~\eqref{#1}}        
\newcommand{\fref}[1]{Fig.~\ref{#1}}   
\newcommand{\cref}[1]{Chap.~\ref{#1}}  
\newcommand{\sref}[1]{Sec.~\ref{#1}}  
\newcommand{\tref}[1]{Table~\ref{#1}}    
\newcommand{\rref}[1]{Ref.~\cite{#1}}

\newcommand{\xz}[1]{\textcolor{lightgray}{#1}}

\usepackage{fontawesome}

\usepackage{titlesec}

\usepackage{hyperref}

\newcommand{\doi}[1]{{doi:~\href{http://doi.org/#1}{#1}}\rmFullStop}

\newcommand*{\rmFullStop}{\rmifnextchar{.}{}{}}

\makeatletter
\newcommand{\rmifnextchar}[3]{%
  \begingroup
  \ltx@LocToksA{\endgroup#2}%
  \ltx@LocToksB{\endgroup#3}%
  \ltx@ifnextchar{#1}{%
    \def\next{\the\ltx@LocToksA}%
    \afterassignment\next
    \let\scratch= %
  }{%
    \the\ltx@LocToksB
  }%
}
\makeatother

\usepackage{colortbl}

\definecolor{light-gray}{gray}{0.6}

\usepackage{algorithm2e}
\usepackage{xcolor}

\SetAlFnt{\footnotesize}

\SetAlCapSty{xAlCapSty}

\definecolor{commentcolor}{HTML}{1E4D2B}

\SetCommentSty{xCommentSty}

\SetNlSty{mynlfont}{}{} 

\LinesNumbered

\SetSideCommentRight

\DontPrintSemicolon

\RestyleAlgo{algoruled}

\usepackage{algorithm2e}
\usepackage{etoolbox}
\usepackage{xstring}
\usepackage{calc}
 
\newlength{\xalgowidth}
\newlength{\xalgoremainder}
\newlength{\xindentwidth}

\newenvironment{vAlgorithm*}[3][]{
  \setlength{\xalgowidth}{#2} 
  \setlength{\xindentwidth}{#3} 
  \setlength{\xalgoremainder}{\textwidth-\xalgowidth} 
  \SetCustomAlgoRuledWidth{\xalgowidth} 
  \IncMargin{\xindentwidth}
  \begin{algorithm*}[#1]
}
{
  \end{algorithm*} 
  \DecMargin{\xindentwidth}
}

{
  \end{algorithm} 
  \DecMargin{\xindentwidth}
}

\makeatletter
\patchcmd{\@algocf@start}{%
\begin{lrbox}{\algocf@algobox}%
}{%
\rule{0.5\xalgoremainder}{\z@}
\begin{lrbox}{\algocf@algobox}%
\begin{minipage}{\xalgowidth}%
}{}{}
\patchcmd{\@algocf@finish}{%
\end{lrbox}%
}{%
\end{minipage}%
\end{lrbox}%
}{}{}
\makeatother

\usepackage{accents}

\definecolor{needcolor}{HTML}{C62828}

\usepackage{amsmath}
\usepackage{bm}
\usepackage{nicematrix}

\definecolor{poscolor}{HTML}{1e88e5}
\definecolor{negcolor}{HTML}{e53935}
\definecolor{offcolor}{HTML}{A93C93}

\newcommand{\CM}[6]{%
\begin{table}[t]
\centering
\caption{#5}
\label{#6}
\vspace{-0.15in}
\scalebox{0.8}{
\begingroup
\setlength{\tabcolsep}{6pt} 
\renewcommand{\arraystretch}{1.2} 
\begin{NiceTabular}{cccc}

& & \Block{1-2}{\textit{Data}} \\
 & & \Block[c,color=poscolor,draw=black,respect-arraystretch]{}{Actually\\ Positive (1)} & \Block[c,color=negcolor,draw=black,respect-arraystretch]{}{Actually\\ Negative (0)} \\
\Block{2-1}{\rotate \textit{Model}} & \Block[c,color=poscolor,draw=black,respect-arraystretch]{}{Predicted\\ Positive (1)} & \Block[c,color=poscolor,draw=black,fill=poscolor!5]{}{#1} & \Block[c,color=offcolor,draw=black,fill=offcolor!5]{}{#2} \\
& \Block[c,color=negcolor,draw=black,respect-arraystretch]{}{Predicted\\ Negative (0)} & \Block[c,color=offcolor,draw=black,fill=offcolor!5,respect-arraystretch]{}{#3} & \Block[c,color=negcolor,draw=black,fill=negcolor!5,respect-arraystretch]{}{#4} \\
\end{NiceTabular}%
\endgroup
}
\end{table}
}

\usepackage[displaymath,switch]{lineno}

\confshortname{IDETC/CIE2023}

\conffullname{Proceedings of the ASME 2023\\  International Design Engineering Technical Conferences and\\ Computers and Information in Engineering Conference}

\confdate{20-23}
\confmonth{August}
\confyear{2023}
\confcity{Boston}
\confcountry{Massachusetts}
\papernum{IDETC2023-114592}

\title{On the Use of Geometric Deep Learning for the Iterative Classification and Down-Selection of Analog Electric Circuits}

\author{}
\author{Anthony Sirico Jr.\thanks{Corresponding author, \texttt{\href{mailto:anthony.sirico@colostate.edu}{anthony.sirico@colostate.edu}}}, Daniel~R.~Herber
\affiliation{
Department of Systems Engineering \\
Colorado State University \\
Fort Collins, CO 80523 \\
Email:~\texttt{\{\href{mailto:anthony.sirico@colostate.edu}{anthony.sirico}, \href{mailto:daniel.herber@colostate.edu}{daniel.herber}\}{@}colostate.edu}
}
}
\author{}

\usepackage{amsmath}
\usepackage{xcolor}
\usepackage{soul}
\sethlcolor{green}

\begin{document}
 \setlength{\parskip}{0pt}
 \setlength{\parsep}{0pt}
 \setlength{\headsep}{0pt}

\setlength{\topsep}{0pt}

\abovedisplayshortskip=3pt
\belowdisplayshortskip=3pt
\abovedisplayskip=3pt
\belowdisplayskip=3pt

\titlespacing*{\section}{0pt}{18pt plus 1pt minus 1pt}{3pt plus 0.5pt minus 0.5pt}

\titlespacing*{\subsection}{0pt}{9pt plus 1pt minus 0.5pt}{1pt plus 0.5pt minus 0.5pt}

\titlespacing*{\subsubsection}{0pt}{9pt plus 1pt minus 0.5pt}{1pt plus 0.5pt minus 0.5pt}

\maketitle

\begin{abstract}\noindent
 
\textit{Many complex engineering systems can be represented in a topological form, such as graphs.
This paper utilizes a machine learning technique called Geometric Deep Learning (GDL) to aid designers with challenging, graph-centric design problems. 
The strategy presented here is to take the graph data and apply GDL to seek the best realizable performing solution effectively and efficiently with lower computational costs. 
This case study used here is the synthesis of analog electrical circuits that attempt to match a specific frequency response within a particular frequency range.
Previous studies utilized an enumeration technique to generate 43,249 unique undirected graphs presenting valid potential circuits.
Unfortunately, determining the sizing and performance of many circuits can be too expensive.
To reduce computational costs with a quantified trade-off in accuracy, the fraction of the circuit graphs and their performance are used as input data to a classification-focused GDL model.
Then, the GDL model can be used to predict the remainder cheaply, thus, aiding decision-makers in the search for the best graph solutions.
The results discussed in this paper show that additional graph-based features are useful, favorable total set classification accuracy of 80\% in using only 10\% of the graphs, and iteratively-built GDL models can further subdivide the graphs into targeted groups with medians significantly closer to the best and containing 88.2 of the top 100 best-performing graphs on average using 25\% of the graphs.}
\end{abstract}

\vspace{1ex}
\noindent Keywords: machine learning, geometric deep learning, graph classification, graph-based design, circuit synthesis

\section{INTRODUCTION}
\label{intro}
Mathematical graphs can be used to represent many systems and decisions because of their ability to capture discrete compositional and relational information. 
For decades, studies have employed different graph representations to capture their respective problems \cite{b5, b6, b37, b95, b96, b97, b102, b105}, and for well over a century, researchers have utilized graph enumeration to understand engineering design problems to help in decision making \cite{b3, b59, b6, b37, b38}. 
Today, many engineering design problems, including the construction of the ``system architecture'', are increasing in scope and complexity to a point where traditional discrete and continuous presentations are insufficient to represent the system \cite{b6}.

The system architecture is a conceptual model capturing the structure, behavior, rules, etc., of a product, process, or element \cite{b18} and is often the foundation on which it is designed, built, and operated. 
Many studies have concentrated on the effective representation of system architectures using graph theory \cite{b29, b91, b92}, where the goal is often a set of useful architectures that are feasible with respect to constraints.
Furthermore, one or more value metrics (e.g.,~performance or cost) might be determined for the alternative architectures so that the designer might sort through the candidates.
One method for generating all these options is graph enumeration, where a complete and ordered listing of the potential graphs is produced for some prescribed structure \cite{b3, b30, b31}.
However, this approach can lead to an enormous amount of potential solutions depending on the problem at hand and a selected representation.
Paired with the increasing complexity of modern systems, the result is an exponential increase in computational costs making the decision-making process for the designer or system architect increasingly challenging.
These challenges drive the need for an approach that facilitates the decision-making for larger and more complex graph-centric design problems.

In this paper, we consider deep learning, specifically \textit{Geometric Deep Learning (GDL)}, as a potential strategy to address these issues.
Deep learning is defined as machine learning models composed of multiple processing layers capable of learning data representations with multiple levels of abstraction \cite{b21}, and
``deep'' refers to a larger number of hidden layers within the neural network. 
Now, GDL is an umbrella term encompassing an emerging technique that generalizes neural networks to Euclidean and non-Euclidean domains, such as graphs, manifolds, meshes, or string representations \cite{b19}, and uses \textit{Graph Neural Networks (GNNs)}.
In essence, GDL encompasses approaches that incorporate information on the input variables' structure space and symmetry properties and leverage it to improve the quality of the data captured by the model. 
GDL has immense potential and is widely used in other scientific communities, including molecular representations \cite{b20, b23, b24}, materials science \cite{b25}, architecture \cite{b28}, and the medical field \cite{b26, b27}.
Within engineering design, there have been applications of GNNs to airfoil design \cite{b106}, structural mechanics using mesh-based physics simulations \cite{b107}, wind-farm
power estimation \cite{Park2019a}, distributed circuit design \cite{b110}, and decision-making processes involved with shared mobility systems \cite{b108}.
Additionally, GNNs have aided in component function classification by training on assembly and flow relationships \cite{b109}.

Most of these engineering problems presented do not have extensive datasets, and perhaps due to this lack of relevant graph-based design datasets, there is limited usage of GDL or other similar graph-based, machine-learning approaches in the engineering design down-select process, despite the availability of the tools and documented success in other fields \cite{b101}.
Many studies utilize machine learning for the representation and selection of objects in 3D space, often in computer-aided design (CAD) applications \cite{b103, b104}.
However, there are some key differences between meshes and other graph representations, including size, structure, locality, and geometric interpretation (or lack thereof).
It is the intent of this study to provide another type of graph design example, and data set \cite{gdlgithub} illustrating how GDL can aid in the engineering design processes. 

Here we are particularly motivated by graph-centric design problems where generating many potential graphs is feasible, but determining the value or performance of each option is too expensive.
Previous work has shown promising results on the specific dataset used in this article but required a significant training set size and time \cite{Guo2019a}.
Here, the proposed approach uses GDL to reduce the computational expense of classifying what might be ``good'' or ``bad'' options with a trade-off in classification accuracy, utilizing much smaller training sets.

The remainder of the paper is organized as follows:
Section~\ref{background} discusses the necessary background for graph design and GDL.
In \sref{method}, we go over the proposed approach, including graph classification, the machine learning model, and the metrics used to determine the performance of the models.
Section~\ref{case} discusses the case study on electric circuit frequency response
matching graph design, and \sref{results} describes several experiments conducted to explore GDL on this problem.
Lastly, we conclude in \sref{conclusion} with the final discussions and future work.

\section{BACKGROUND}
\label{background}

To better understand the motivation and techniques behind the proposed methodology, this section will highlight additional key aspects of graph theory, system representation through graphs, GDL, and GNNs.

\subsection{Graph Theory}
\label{gt}
A \textit{graph} $G$ is a pair of sets $(V,E)$ where $E \subseteq [V]^2$ (i.e., $E$ is a two-element subset of $V$). $E$ represents the edges of the graph, while $V$ is the set of its vertices or nodes \cite{b7}.
A graph has various properties.
For example, a graph's order, denoted by $n = |G|$, is the number of vertices in the graph.
Each of those vertices has an associated degree value, which is the number of neighbors or edges to a vertex and is denoted by $d_G(v) = d(v)$.

While there are different ways to represent a graph mathematically, here we focus on one known as the \textit{adjacency matrix} $\textbf{A} = (a_{i,j})_{n \times n}$ and is defined as:
\begin{align}
a_{i,j} \coloneqq 
\begin{cases}
1, & \text{ if } (v_i,v_j) \in E \\
0, & \text{ otherwise}
\end{cases}
\label{adjeq}
\end{align}

\noindent For simple, undirected graphs, the matrix is symmetric and the diagonal of $\textbf{A}$ will be all zeros, which indicates that the graph has no self-loops.

Another important concept is \textit{graph isomorphism}.
Say we have two graphs, $G_1 = (V_1,E_1)$ and $G_2 = (V_2,E_2)$.
These two graphs are isomorphic, denoted $G1 \cong G_2$, if there is a bijection, $\varphi$, from $V_1 \rightarrow V_2$ such that $(v_i,v_j) \in E_1 \leftrightarrow (\varphi(v_i),\varphi(v_j))\in E_2$ for all $v_i,v_j \in V$ \cite{b78}.
We also consider a feature or \textit{labeled graph isomorphism}.
Here we consider the same two graphs from the previous definition, but they contain a third feature.
Specifically, we will consider this feature as a vertex label, denoted $X$, where $G_1 = (V_1,E_1,X_1)$ and $G_2 = (V_2,E_2,X_2)$.
The graphs will be isomorphic as long as the vertex label property is preserved under some valid bijection $\varphi$.
More generally, a features matrix $\textbf{X} \in \mathbb{R}^{n \times c}$ can have as many columns as necessary to represent the $c$ features associated with each vertex.

\subsubsection{Representing Electric Circuits as Graphs.}

\begin{figure}[t]
\centering
\includegraphics[scale=0.95]{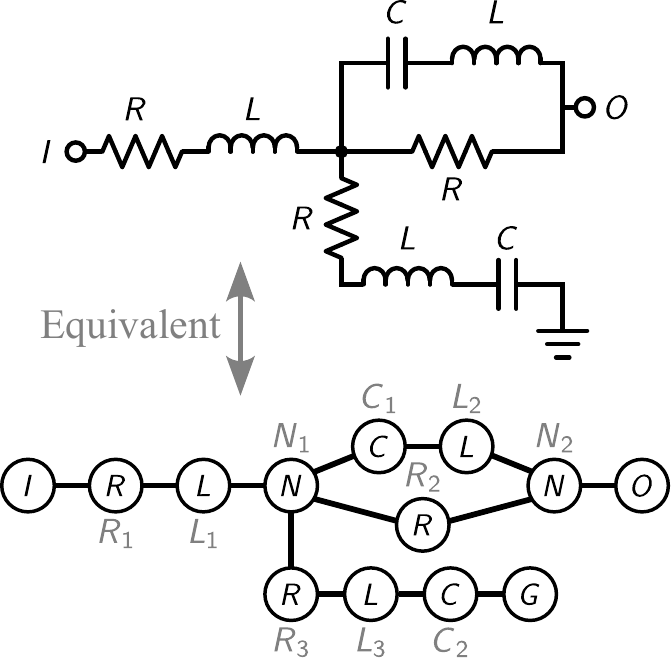}
\caption{Electrical circuit schematic represented as a vertex-labeled graph.}
\label{circgraph}
\end{figure}

The concepts from the previous section can be used to represent engineering systems.
The particular type of system used in the case study here is electric $RLC$ circuits, so they will be used to illustrate their representation as a graph.
Looking at the left side of \fref{circgraph}, we can see a basic $RLC$ circuit schematic.
Then on the right side, we have the same circuit represented pictorially as a graph.
Each vertex in the graph is now labeled as the corresponding component where $I$ and $O$ represent the input and output nodes, $R$ is a resistor, $L$ is an inductor, $C$ is a capacitor, and lastly, $N$ represents a voltage node with constant voltage. Also, other graph representations of the same circuit are possible.

For the graph in \fref{circgraph}, an adjacency matrix $\mathbf{A}$ representing its structure and vertex list $\mathbf{V}$ representing the vertex labels are:
\begin{align}
\begingroup 
\setlength\arraycolsep{3pt}
\setcounter{MaxMatrixCols}{20}
\textbf{A} = \begin{bmatrix}  
\xz{0} & 1 & \xz{0} & \xz{0} & \xz{0} & \xz{0} & \xz{0} & \xz{0} & \xz{0} & \xz{0} & \xz{0} & \xz{0} & \xz{0}\\ 1 & \xz{0} & 1 & \xz{0} & \xz{0} & \xz{0} & \xz{0} & \xz{0} & \xz{0} & \xz{0} & \xz{0} & \xz{0} & \xz{0}\\ \xz{0} & 1 & \xz{0} & 1 & \xz{0} & \xz{0} & \xz{0} & \xz{0} & \xz{0} & \xz{0} & \xz{0} & \xz{0} & \xz{0}\\ \xz{0} & \xz{0} & 1 & \xz{0} & 1 & \xz{0} & \xz{0} & \xz{0} & 1 & 1 & \xz{0} & \xz{0} & \xz{0}\\ \xz{0} & \xz{0} & \xz{0} & 1 & \xz{0} & 1 & \xz{0} & \xz{0} & \xz{0} & \xz{0} & \xz{0} & \xz{0} & \xz{0}\\ \xz{0} & \xz{0} & \xz{0} & \xz{0} & 1 & \xz{0} & 1 & \xz{0} & \xz{0} & \xz{0} & \xz{0} & \xz{0} & \xz{0}\\ \xz{0} & \xz{0} & \xz{0} & \xz{0} & \xz{0} & 1 & \xz{0} & 1 & 1 & \xz{0} & \xz{0} & \xz{0} & \xz{0}\\ \xz{0} & \xz{0} & \xz{0} & \xz{0} & \xz{0} & \xz{0} & 1 & \xz{0} & \xz{0} & \xz{0} & \xz{0} & \xz{0} & \xz{0}\\ \xz{0} & \xz{0} & \xz{0} & 1 & \xz{0} & \xz{0} & 1 & \xz{0} & \xz{0} & \xz{0} & \xz{0} & \xz{0} & \xz{0}\\ \xz{0} & \xz{0} & \xz{0} & 1 & \xz{0} & \xz{0} & \xz{0} & \xz{0} & \xz{0} & \xz{0} & 1 & \xz{0} & \xz{0}\\ \xz{0} & \xz{0} & \xz{0} & \xz{0} & \xz{0} & \xz{0} & \xz{0} & \xz{0} & \xz{0} & 1 & \xz{0} & 1 & \xz{0}\\ \xz{0} & \xz{0} & \xz{0} & \xz{0} & \xz{0} & \xz{0} & \xz{0} & \xz{0} & \xz{0} & \xz{0} & 1 & \xz{0} & 1\\ \xz{0} & \xz{0} & \xz{0} & \xz{0} & \xz{0} & \xz{0} & \xz{0} & \xz{0} & \xz{0} & \xz{0} & \xz{0} & 1 & \xz{0} \end{bmatrix}
\qquad 
\textbf{X} = \begin{bmatrix} I\\R\\L\\N\\C\\R\\L\\N\\O\\R\\L\\C\\G\end{bmatrix}
\label{adjmat}
\endgroup
\end{align}

\noindent where $1$ represents a connection between two vertices.

\subsection{Graph Enumeration and Optimization}
\label{enumeration}

Designers and system architects still often rely on engineering intuition and trial-and-error methods when exploring graph-centric problems.
However, these approaches can lead to a lack of novel and satisfactory solutions within time restrictions.

A potential systematic strategy is \textit{graph enumeration}, which are techniques for generating (or sometimes simply counting) nonisomorphic graphs with particular properties \cite{b89}.
The properties can be quite diverse and are sometimes termed network structure constraints \cite{b31}.
Some examples include bounding the potential degrees of the vertices in the graph or a maximum cost associated with the graph \cite{b59}.
It is important to note that the number of valid graphs is usually combinatorial in nature; thus, the computational costs for just generating an enumeration can become quite expensive.
However, two different graph enumeration algorithms might produce the desired enumeration, but one might do it more efficiently \cite{b59}.

In any case, there are several important engineering applications where the enumeration of graphs, often with thousands to millions of entries, is practical. 
Even if graph enumeration is not possible, generating a partial listing (either algorithmically or manually) may still be desirable to explore potential solutions.
In this work, it is assumed that there is a generated set of graphs, denoted $\mathcal{G}$, that contains the graphs of interest.
For the purposes of the case study, this is an enumeration of $RLC$ circuit graphs with certain properties based on \cite{b5}.

While having a list of novel graphs can be useful on its own, in many engineering domains, the graph is only an intermediary representation used to compute one or more value metrics of interest.
For example, the graph might correspond to a physics-based system model that is used to determine the comfort and handling of an automotive suspension \cite{b94}.
For this work, we will consider a single performance metric, denoted $J(G_i)$, that is a function of the graph, and the natural goal is seeking graphs that minimize this ``performance'' metric:
\begin{align}
\underset{G_i}{\textrm{minimize:}} \quad & J(G_i)
\label{archopt}
\end{align}

One of the challenges in engineering design with graphs is sometimes the computational cost of $J(G_i)$ is quite expensive, perhaps many orders of magnitude larger than generating the graph $G_i$ itself.
The source of this cost can be quite diverse, including high-fidelity simulations, optimization, human-centric evaluation, and physical experiments.
The focus of this work is when this cost is prohibitive. 

Based on the classification put forth by \cite{b93}, we define the following three types of graph-centric design problems:
\begin{enumerate}[nolistsep,label=$\bullet$]
   
\item \textit{Type 0} --- All desired graphs can be generated and so can their performance metric $J(G_i)$ within time $T$

\item \textit{Type 1} --- All desired graphs can be generated, but only some of the performance metrics $J(G_i)$ can be evaluated within time $T$; the performance assessment is too expensive

\item \textit{Type 2} --- All desired graphs cannot be generated within time $T$
    
\end{enumerate}

\noindent where $T$ is the amount of time allocated to complete the graph design study. 
This work focuses on methods for Type 1 problems (using data from a large Type 0 study).

\subsection{Geometric Deep Learning}
\label{gdl}

Deep learning models have been very successful when training on images \cite{b32}, text \cite{b33}, and speech \cite{b34}, but these contain an underlying Euclidean structure. 
Graphs are fundamentally different from the more common Euclidean data used in most deep learning applications (language, images, and videos).
Euclidean data has an underlying grid-like structure.
For example, an image can be translated to an $(x,y,z)$ Cartesian coordinate system, where each pixel is located at an $(x,y)$ coordinate, and $z$ represents the color. 
But, this approach does not work for all problems because certain operations require many dimensions.
This ``flat'' representation has limitations, such as not being able to represent hierarchies and other direct relationships. 

This motivated the study of hyperbolic space for graph representation, or non-Euclidean learning \cite{b72, b73}, which could potentially perform those same operations more flexibly, and GDL methods that can take advantage of the rich structure of such data to learn better representations and make better predictions \cite{b19}.
For example, many real-world data sets are naturally structured as graphs, where the \textit{relationships} between data points are more critical than individual data points. 
Euclidean-based methods struggle with such data, as they are designed to operate on flat, unstructured data. 
On the other hand, GDL methods can directly exploit the structure of the data, leading to better outcomes. 

Since its inception, GDL can be applied to all forms of data represented as geometric priors \cite{b1}. 
Geometric priors encode information about the geometry of the data, such as smoothness, the sparsity of the data, or the relationship between the data and other variables, allowing us to work with data of higher dimensionality.
Symmetry is essential in GDL and is often described as invariance and equivariance. 
Invariance is a property of particular mathematical objects that remain unchanged under certain transformations, while equivariance is a property of certain relations whereby they remain in the same relative position to one another under certain transformations \cite{b35, b36}.
For example, consider the graph isomorphism property from Sec.~\ref{gt}.
Another advantage of geometric deep learning is its flexibility toward a broader range of data types and problems.
This flexibility makes it a powerful tool for solving real-world problems that Euclidean methods cannot.

\subsection{Graph Neural Networks} 
\label{gnn}
Similar to traditional deep learning models that use Convolutional Neural Networks (CNNs), GDL utilizes GNNs, which are used to learn representations of graph-structured data, such as social networks, molecules, and computer programs. 
GNNs are very similar to CNNs in that they can extract localized features and compose them to construct representations \cite{b43}. 
As previously mentioned, the critical difference between the two is that GNNs can learn on non-euclidean data and handle data that is not evenly structured, like images or text. 
Lastly, GNNs can learn from data that is not labeled, known as unsupervised learning. 
This is important because many real-world datasets are not labeled.

Training a GNN for graph classification (i.e., determining if a graph has a characteristic or not) is relatively simple, and there is typically a three-step process that needs to occur: 1) embed each node by performing multiple rounds of message passing, 2) aggregate node embeddings into a unified graph embedding, and then 3) train a final classifier on the graph embedding.
We take a closer look at how this occurs in \sref{model}, where we discuss each layer of the GNN used here.

\section{METHODOLOGY}
\label{method}

The methodology for apply GDL on graph-based design problems will have a graph classification approach.
This section will discuss this concept along with the selected model, hyperparameters, and metrics used to evaluate performance of the model.

\subsection{Graph Classification}
\label{gclass}

\begin{figure}[t]
\centering
\includegraphics[scale=0.95]{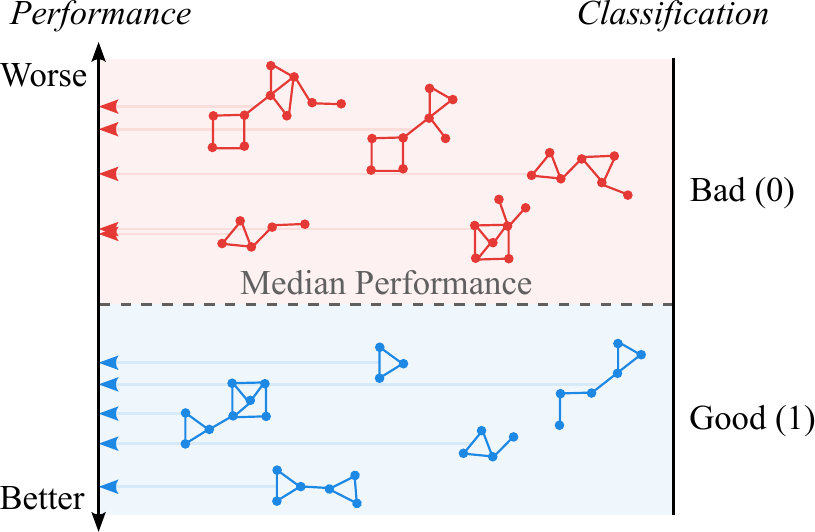}
\caption{Illustration of graph classification based on performance values.}
\label{fig:graphclass}
\end{figure}

\textit{Graph classification} refers to classifying graphs based on selected labels.
Here, we consider \textit{supervised} graph classification, where we have a collection of graphs $\mathcal{G}$, each with its label $J(G_i)$, used to train a model using additional graph properties, such as the structure, embeddings, and node data.
These features help the model discriminate between the graphs, improving the prediction of the labels.
An illustration of graph classification is shown in \fref{fig:graphclass}.
Here, the median performance value might be used as the determination point to divide our graphs into binary classes representing ``good'' graphs in the lower half of the observed performance values and ``bad'' graphs in the upper half. 

The rationale for this approach over regression was to better reflect the designer's intention in early-stage conceptual design (i.e., flexible down-selecting).
Often the goal is not to narrow the potential graphs down to \textit{one} particular graph, but rather a \textit{group} of ``good'' or promising graphs that would be analyzed further (often at a higher fidelity due to assumptions made in modeling and other areas during graph design).
Furthermore, the performance values $J(G_i)$ in the case study have significant variations and many coarse values.
Therefore, the intent is to show the ``belongingness'' of the graphs to a specific group versus assigning specific values to graphs.

There are various methods of graph classification, including Deep Graph Convolutional Neural Networks (DGCNN) \cite{b2}, hidden layer representation that encodes the graph structure and node features \cite{b8}, EigenPooling \cite{b12}, and differentiable pooling \cite{b81}.
It has been used in many different studies, such as learning molecular fingerprints \cite{b80}, text categorization \cite{b98}, encrypted traffic analysis \cite{b99}, and cancer research \cite{b100}.

\subsection{Datasets}
\label{sec:datasets}

Based on the Type 1 problem classification considered here from \sref{enumeration}, we will consider the case when only some of the performance values $J(G_i)$ for $G_i \subset \mathcal{G}$ are known.
This will divide the graphs into two sets as follows:
\begin{align}
\mathcal{G} \equiv \mathcal{G}_{all} = \mathcal{G}_{known} \cup \mathcal{G}_{unknown}
\end{align}

\noindent where $\mathcal{G}_{known}$  is the set of graphs with known values for $J(G_i)$ and $\mathcal{G}_{unknown}$ represents graphs with unknown $J(G_i)$ values (and this is what the GDL model is for).
We will denote the sizes of the sets as $|\mathcal{G}_{all}| = N_{all}$, $|\mathcal{G}_{known}| = N_{known}$, and $|\mathcal{G}_{unknown}| = N_{unknown}$. They also satisfy:
\begin{align}\label{eq:N}
N_{all} = N_{known} + N_{unknown}
\end{align}

\noindent We also note that this is a bit different than other types of datasets used in machine learning as there is a known, finite amount of potential inputs.
The goal here is to develop accurate models where $N_{known} \ll N_{unknown}$.

As is typical in machine learning, we will create two subsets from $\mathcal{G}_{known}$ :
\begin{align}
\mathcal{G}_{known} = \mathcal{G}_{training} \cup \mathcal{G}_{validation}
\end{align}

\noindent where $\mathcal{G}_{training}$ is the training dataset and $\mathcal{G}_{validation}$ is the validation dataset.
The model fits using $\mathcal{G}_{training}$, and the fitted model is used to predict the responses for the observations in $\mathcal{G}_{validation}$ \cite{b46}. 
For example, after each iteration, the model will adjust its weights accordingly and test them on the validation set, which can help understand model performance and adjust options as necessary. 
The unknown set is used when the model is finalized, and you want to now test it on data that the model has not seen.

\begin{figure}[t]
\centering
\includegraphics[scale=0.95]{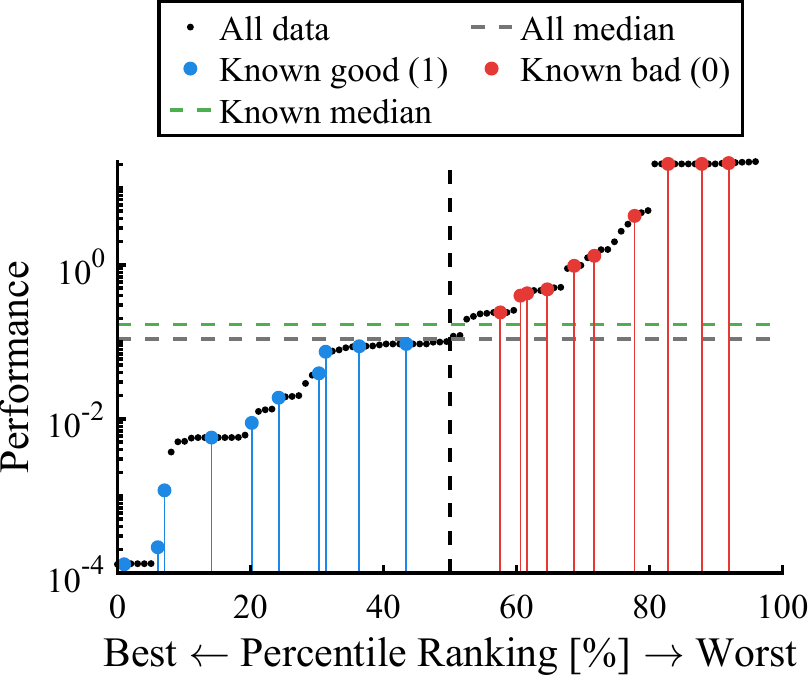}
\caption{Illustration of $\mathcal{G}_{all}$ and $\mathcal{G}_{known}$ for 100 graphs, the performance-based classification of $\mathcal{G}_{known}$, and the potential difference between the medians.}
\label{perc}
\end{figure}

As shown in \fref{fig:graphclass}, the binary classification of graphs in $\mathcal{G}_{known}$  is based on the median performance value (or potentially some other dividing line using the known data).
Since we only know $J(G_i)$ in $\mathcal{G}_{known}$  rather than $\mathcal{G}_{all}$, the median performance value used for the initial labeling might be different than the true value if we had all the performance values for $\mathcal{G}_{all}$.
However, as the relative size of $\mathcal{G}_{known}$  increases, this error will become small.
The concepts in this section are illustrated in \fref{perc}.
Here we have 100 graphs in $\mathcal{G}_{all}$ and 20 randomly selected graphs for $\mathcal{G}_{known}$.
The true median and $\mathcal{G}_{known}$-based median do differ, and two graphs are between these lines.

\subsection{The Model}
\label{model}

There are five layers for the model used in this study: three graph convolutional layers \cite{b65}, a mean pool layer, and a linear layer for the final readout.
To accomplish the graph classification goal, the model uses: 1) the convolutional layers to embed each node through message passing, 2) agglomerating the node embeddings to create a graph embedding, then 3) use the graph embedding to train the classifying layer.

\subsubsection{Graph Convolutional Layer.}~The Graph Convolutional Layer (GCN) layers determine the output features $\textbf{X}'$ by:
\begin{align}
\textbf{X}'_i = \textbf{W}_1\textbf{X}_i+\textbf{W}_2 \sum_{j\in \mathrm{N}(i)}e_{j,i}\cdot \textbf{X}_j
\label{gcn}
\end{align}

\noindent where $\textbf{X}$ is the input feature of each node (as discussed in \sref{gt}), $(\textbf{W}_1,\textbf{W}_2)$ represents the weights adjusted after each iteration while training, and $e_{j,i}$ is the edge weight from source node $j$ to target node $i$ \cite{b65}.
Each GCN layer also uses the ReLU activation function:
\begin{align}
f(x) = \max(0,x)
\label{relu}
\end{align}

\noindent which returns 0 if it receives any negative input, but for any positive value $x$, it returns that value.
Finally, the outcome of \eref{gcn} is passed through the activation function:%
\begin{align}
\textbf{X}_{i+1} = f(\textbf{X}_i) 
\label{gcn2}
\end{align}

\noindent which helps us obtain localized node embeddings.

\begin{figure}[t]
\centering
\includegraphics[width=0.90\columnwidth]{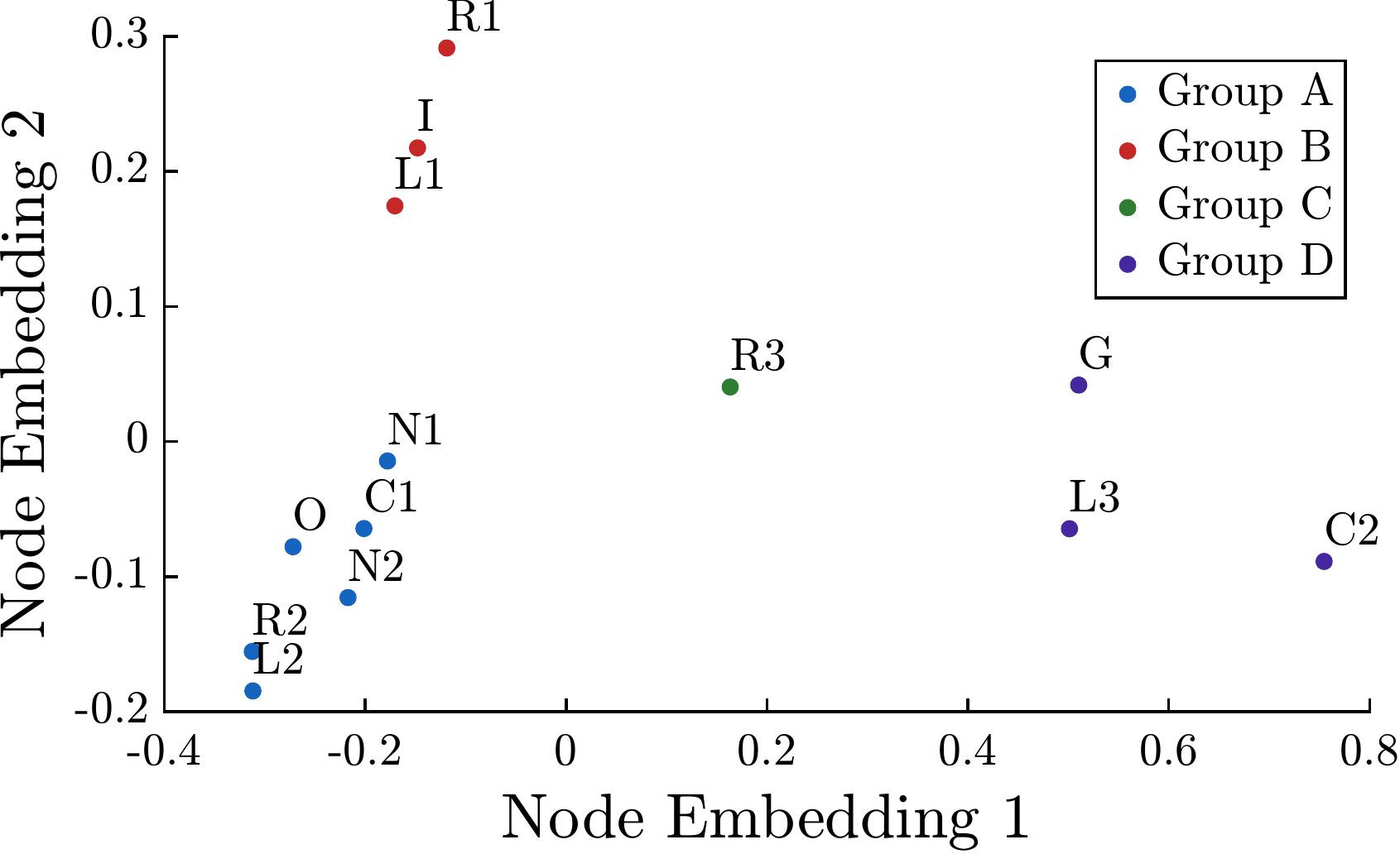}
\caption{Calculated node embeddings for the graph in \fref{circgraph}.}
\label{embed}
\end{figure}

Node embeddings are a mapping of a graph's nodes or vertices into an $N$-dimensional numerical space.
This process creates numerical features as vectors from the graph structure. 
Nodes similar to one another will be spaced close together in what are known as communities.
As an example, we can use Node2Vec \cite{b79} to compute the node embeddings, visualized in \fref{embed} using the graph from \fref{circgraph}.
Comparing the two figures, we can see how Node2Vec analyzes the input circuit and interpret related nodes through the embeddings.
In \fref{circgraph}, the three nodes ($I_1$,$R_1$,$L_1$) are shown in a grouping in \fref{embed}, as well as the parallel components community ($N_1$,$C_1$,$L_2$,$R_2$,$N_2$,$O$); this is consistent since these components are connected and near one another.

\subsubsection{Global Mean Pooling Layer.}~The next layer is \textit{Global Mean Pooling}, which takes in the last output from the GCN layers and returns an output $\textbf{r}$ by averaging the node features $\textbf{X}$ to include node embeddings for each graph across all nodes $N$, creating a graph embedding:
\begin{align}
\textbf{r}_i = \frac{1}{N_i}\sum^{N_i}_{n=1} \textbf{X}_n
\label{gmp}
\end{align}

\subsubsection{Linear Layer.}~The final layer is the classification layer, which takes the mean pooling layer output as its input and applies the following linear transformation:
\begin{align}
\mathbf{y} = \mathbf{r} \mathbf{A}^T + \mathbf{b}
\label{lin}
\end{align}
However, before this layer, a \textit{dropout} is applied, which randomly zeros some of the elements of the input tensor by some assigned probability using samples from a Bernoulli distribution \cite{b82}.

\subsection{Model Hyperparameters and Options}
\label{hyp}

For the selected model architecture, there are several items that can be tuned so that the model trains well on the type of data provided.
For more information on hyperparameters and tuning, please refer to the following sources \cite{b44, b45, b46}.

\subsubsection{Learning Rate.}~The model's learning rate ($LR$) is a positive scalar value that determines the size of the step response to the estimated error as the weights are updated with each epoch and is considered one of the more important hyperparameters \cite{b45}.
Typically, this value ranges from $0.0$ to $1.0$. 
Choosing a value too small can result in a longer training time, and choosing a value too large can result in an unstable training process.
For these case studies, we use a $LR$ of $0.001$.

\subsubsection{Number of Epochs.}~An epoch refers to the number of training iterations through the entire training dataset \cite{b83}.
The data is passed through the model during each epoch, updating the weights.
Epochs can often range from hundreds to thousands, allowing the model to train until the error in the model is minimized.
The number of epochs is explored in the different experiments in the case study. 

\subsubsection{Optimization Algorithm.}~An optimization algorithm is a procedure for finding the input parameters for a function that results in the minimum or maximum of the function \cite{b83}.
In this model, we chose to use the Adaptive Movement Estimation (ADAM) algorithm, a stochastic optimization method that computes individual learning rates for different parameters of the first and second moments of the gradients \cite{b84}.

\subsubsection{Loss Function.}~A loss function computes the distance between the current output of the model with the expected out of the model.
Then, this metric measures how well the model is performing.
There are various loss functions, and the one used in this model is the \textit{cross-entropy} loss function, typically used for classification problems.
It measures the difference between two probability distributions for a given variable.

\subsubsection{Batch Size.}~While training, we can decide on batch size or the number of data points to work through before updating the model weights. 
This technique is known as mini-batching and is highly advantageous when training a deep-learning model by allowing the model to scale better to large amounts of data. 
Instead of processing data points one-by-one or all at once, a mini-batch groups a set of data points of intermediary size. 
This hyperparameter will be explored in the case study.

\subsection{The Metrics}
\label{metrics}
This section discusses the metrics used to evaluate the models' performance.

\subsubsection{The Confusion Matrix.}~A confusion matrix (CM), as seen in \fref{excm}, is a two-dimensional matrix where each column contains the samples of the classifier (model) output, and each row contains the sample in the true class (data) \cite{b85}.
For a binary classifier, the top left box represents the \textit{True Positives} ($T_P$), the data points that were classified correctly as ones.
The top right represents the \textit{False Positives} ($F_P$), the points that were classified as ones but are, in fact, zeros.
The bottom left contains the \textit{False Negatives} ($F_N$), the points that were classified as zeros but are actually ones.
Lastly, the bottom right is the \textit{True Negatives} ($T_N$), the points that were correctly classified as zeros.
The values in this matrix are used for many of the following metrics.

\CM{2,050 ($T_P$)}{450 ($F_P$)}{250 ($F_N$)}{2,250 ($T_N$)}{An example confusion matrix with sample values.}{excm}

\subsubsection{Accuracy.}\label{accuracy}~Accuracy is the number of correct predictions divided by the total number of predictions, see \eref{acceq}.
\begin{align}
\textit{Accuracy} = \frac{T_P+T_N}{N}
\label{acceq}
\end{align}

Using the data from \tref{excm}, we can calculate that our classifier had an accuracy of $0.86$.

\subsubsection{Precision}~This metric, also called Positive Predictive Value (PPV), tells us what proportion of positive classifications were actually correct.
This metric is particularly useful when your data has a class imbalance, e.g., more zeros than ones.
When training on a dataset with more of one class than the other, you risk your model classifying all the data as the most frequent class, thus a ``high accuracy''.
That is where \textit{Precision} comes in because it is a class-specific metric defined as:
\begin{align}
\textit{Precision} = \frac{T_P}{T_P+F_P}
\label{ppv}
\end{align}
Once again, using our example, our classifier has a \textit{Precision} of 0.82, which means when it predicts a data point as one, it is correct 82\% of the time.
This same metric can also be applied to the negative value, but for this paper, we will use \textit{Precision} since our case studies focus on finding the \textit{best} architecture.

\subsubsection{Recall.}~This metric is calculated with:
\begin{align}
\textit{Recall} = \frac{T_P}{T_P+F_N}
\label{rec}
\end{align}

\noindent which indicates the proportion of actual positive classifications were identified correctly.
Using the data from \tref{excm}, \textit{Recall} is $0.89$, which states  the model correctly identified $89\%$ of all ones.

\subsubsection{F1 Score.}~This is a combination of both precision and recall into a single metric by calculating the harmonic mean between the two values as:
\begin{align}
\textit{F1 Score} = 2 \cdot \frac{\textit{Precision} \cdot \textit{Recall}}{\textit{Precision} + \textit{Recall}}
\label{f1}
\end{align}
A model with a high \textit{F1 Score} means that both \textit{Precision} and \textit{Recall} were high. 
Continuing with the example, the \textit{F1 Score} for this model is 0.85. 
    
\subsubsection{Matthews Correlation Coefficient (MCC).}~Of the metrics used to evaluate the performance of the classification models, this metric is one of the most important.
Once again, using the values from the confusion matrix, the $MCC$ produces a high score if the model obtained good results in \textit{all} boxes of the confusion matrix \cite{b67, b68}; this means high $T_P$ and $T_N$ values and low $F_P$ and $F_N$ values.
MCC is computed with:
\begin{align}
MCC = \frac{T_P \cdot T_N - F_P \cdot F_N}{\sqrt{(T_P + F_P)(T_P+F_N)(T_N+F_P)(T_N+F_N)}}
\label{mcc}
\end{align}

\noindent $MCC$ ranges from $-1$ to $1$, where $1$ indicates the model can make predictions perfectly, $-1$ indicates that every prediction was incorrect, and $0$ indicates that the model is just as good as random chance.

Finally, in our example, the achieved $MCC$ for this model is $0.72$, which is generally considered a good model.

\subsubsection{Total Set Accuracy.}~When the datasets are broken down into their respective subsets $(\mathcal{G}_{known},\mathcal{G}_{unknown})$, we still want to know how well the predictions are when compared to the $\mathcal{G}_{all}$, if it is available.
Therefore, we define \textit{Total Set Accuracy} as:
\begin{align}
\textit{Total Set Accuracy} = \frac{T_P^{(u)} + T_N^{(u)} + T_P^{(k)} + T_N^{(k)} }{N_{all}}
\label{eq:total-set-acc}
\end{align}

\noindent where $(T_P^{(u)},T_N^{(u)})$ are determined on $\mathcal{G}_{unknown}$ using the trained model, and $(T_P^{(k)},T_N^{(k)})$ are determined using $\mathcal{G}_{known}$, which might not be perfectly correct due to the median difference described in \sref{sec:datasets}.
If there is no misclassification based on the different medians, then $T_P^{(k)} + T_N^{(k)} = N_{known}$.
This metric captures the outcome of a potential real-world scenario where a designer uses both the known and unknown data to classify all graphs.


\section{CASE STUDY: ELECTRIC CIRCUIT FREQUENCY RESPONSE GRAPH-DESIGN PROBLEM}
\label{case}
To explore the capabilities of GDL in graph-based engineering design problems, we utilize the results of a frequency response matching study \cite{b5}.
The dataset and code to replicate these studies are available at Ref.~\cite{gdlgithub}.

\subsection{Enumeration of Circuit Graphs}

Automating the design synthesis of analog circuits has long been a problem of interest with numerous attempts \cite{501647, 4208908, grimbleby}.
Still, these approaches can often fail to meet the needs when encountering unfamiliar and complex design problems.
Even with the addition of heuristics \cite{1083985} and knowledge bases \cite{44506}, determining global solutions, sets of alternatives, and patterns is not straightforward, which leads us to consider enumeration.

As discussed in \sref{enumeration}, we will consider the graph enumeration of electric circuits from Ref.~\cite{b5}.
In particular, the set $\mathcal{G}$ with 43,249 undirected graphs includes all topologies with up to 6 impedance subcircuits with $RC$ components and a required connection to the ground \cite{b5}.
Two examples are shown in Fig.~\ref{fig:pvc} with $RC$ values optimally selected for the design problem presented in the next section.

Many researchers have utilized the enumeration technique of circuit problems \cite{b38, b48, b49} and in other problem areas where the data can be represented as graphs and other enumerable objects \cite{b51, b52, b53}. 
As mentioned, each circuit here is represented by a vertex-labeled graph, meaning every vertex has an associated label representing some circuit concept.
The size of each graph varied between 6 and 20 nodes, all of which had guaranteed different topologies (i.e., no labeled graph isomorphisms from \sref{gt}), which is relatively small for GDL.

\subsection{Frequency Response Matching Optimization}

\begin{figure}[t]
    \centering
    \includegraphics[scale=0.95]{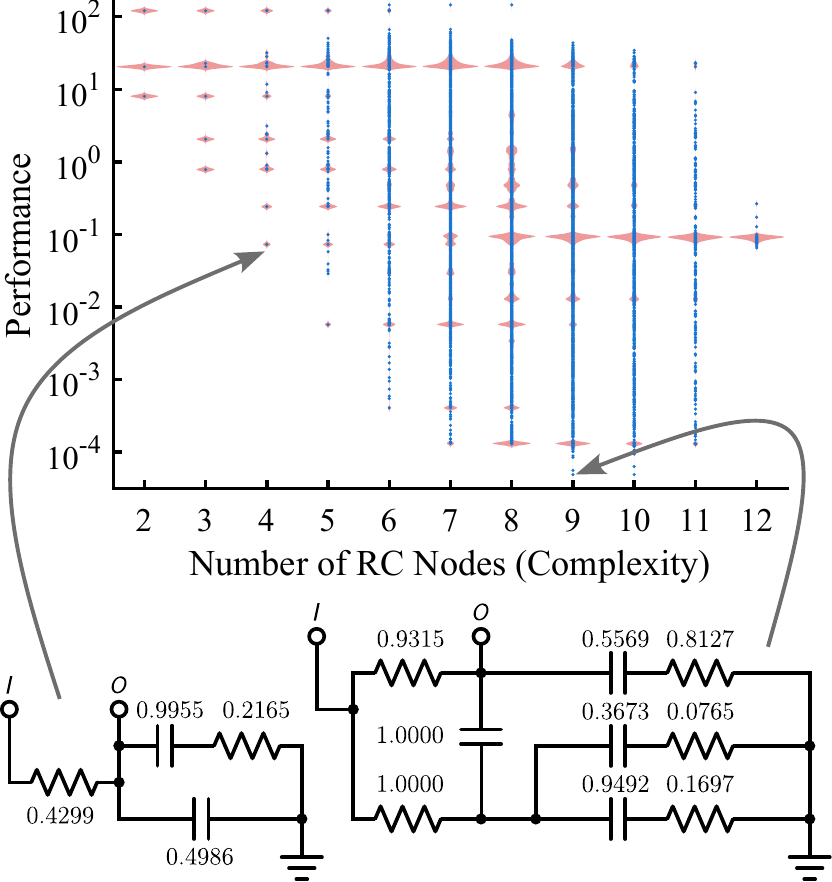}
    \caption{Summary of the number of $RC$ nodes (a complexity metric) versus performance $J(G_i)$ for all 43,249 graphs with individual points and density in the frequency response matching optimization case study dataset.}
    \label{fig:pvc}
\end{figure}

The performance value $J(G_i)$ here is the error between the desired frequency response and the one a selected circuit provides, based originally on the study in \rref{501647}.
A circuit graph $G_i$ does not have an intrinsic value for $J$, rather it is a function of the \textit{tunable values} for the $RLC$ coefficients in the given graph and a selected \textit{optimization problem} with an objective and constraints.
From Ref.~\cite{b5}, we consider the ``set 1'' problem defined as:
\begin{subequations}\label{eq:cir-opt}
\begin{align}
\underset{\mathbf{z}_i = \{\mathbf{R}_i, \mathbf{C}_i\} }{\text{minimize:}} \quad & J = \sum_{k} \left( \log|H_i(j\omega_k,\mathbf{z}_i)| - \log|F(j\omega_k)|  \right)^2 \\
\text{subject to:} \quad & 10^{-2} \leq R_j \leq 10^{0} \quad \text{ for all $R_j$ in $G_i$} \\
& 10^{-2} \leq C_j \leq 10^{0} \quad \text{ for all $C_j$ in $G_i$} \\
\text{where:} \quad & |F(j \omega)| = \sqrt{\frac{2\pi}{10\omega}} \quad 0.2 \leq \frac{\omega}{2\pi}\leq 5
\end{align}
\end{subequations}

\noindent where $|F(j \omega)|$ is the desired frequency response, $H_i(j\omega_k,\mathbf{z}_i)$ is the frequency response for circuit graph $G_i$, $\mathbf{z}_i$ is the collection of optimization variables for the resistors $R$ and capacitors $C$ in graph $G_i$, and $\omega$ is sampled at 500 logarithmically-spaced evaluation points.

Solving this nonlinear constrained least squares optimization problem can be expensive; the original study required over 8 hours to optimize each graph to determine $J(G_i)$.
Therefore, it is desirable to reduce the computational costs to discern good and bad circuit graphs.

The performance values here have a large range, between $4 \times 10^{-4}$ to $2 \times 10^{2}$, as illustrated in Fig.~\ref{fig:pvc}.
Classification is done based on a sampled $\mathcal{G}_{known}$ using the median value of the known performance values, as discussed in \sref{sec:datasets}.
As we have the performance values for all graphs, this problem can serve an a good example to explore the potential effectiveness of GDL in these kinds of problems, so we can compare to classification using $\mathcal{G}_{all}$.

\section{CASE STUDY EXPERIMENTS AND RESULTS}
\label{results}
In this section, we describe the experiments conducted on the engineering graph-design problem dataset from \sref{case}.
The tools and computing architecture are described in App.~\ref{appa} \cite{gdlgithub}.

\subsection{Experiment 1: Establish a Baseline}
\label{exp1r}
In the first experiment, we start by creating a baseline model for future comparisons.
Here we consider a 72\% (31,139 graphs) in the training set $\mathcal{G}_{training}$, 18\% (7,784) in the validation set $\mathcal{G}_{validation}$, and the remaining 10\% (4,324) in  $\mathcal{G}_{unknown}$.
These are by no means recommended distributions, but rather a scenario with lots of data available for the GDL model that will be used to explore what is possible and what might be done to balance accuracy versus efficiency.

The CM for this experiment is in \tref{baselinecm} using the 4,324 graphs in $\mathcal{G}_{unknown}$ that we have not used in any way to train the model (and in practice, you would not have this CM). 
Using \eref{acceq}, we find a {79.1}\% accuracy, fairly close to the baseline models' final accuracy value for $\mathcal{G}_{known}$ of {79.5}\%.
Next, using \eref{ppv}, we calculate that the baseline model has a \textit{Precision} of {77}\%.
Using \eref{rec}, we calculate the baseline models' \textit{Recall} to be {79.9}\%, and lastly, using \eref{f1} and \eref{mcc}, we calculate the \textit{F1 Score} to be {78.6}\% and \textit{MCC} to be {58}\%, respectively.

\subsection{Experiment 2: Additional Graph-based Features}
\label{exp2r}

Here we consider adding additional graph-based features to $\mathbf{X}$ beyond the current vertex labels of $R$, $C$, $G$, etc. where the additional features should have a greater ability to explain the variance in the training data (known as feature engineering and feature selection).
The theory is that these features will improve the model's performance for the same number of training epochs.
The two features added here are eigenvector centrality and betweenness centrality.

Eigenvector centrality computes a node's centrality based on its neighbors' centrality and is a measure of the influence of a node in a graph where a high eigenvector centrality score implies that a node is connected to many nodes that themselves have high scores.
The eigenvector centrality for node $v$ is the $i$-th normalized element of the vector $\mathbf{v}$ from:
\begin{align}
\mathbf{A}\mathbf{v}=\lambda \mathbf{v}
\label{evc}
\end{align}
\noindent where $\mathbf{A}$ is the adjacency matrix and $\lambda$ is the largest eigenvalue \cite{b60}.

Now, the betweenness centrality of a node is the sum of the fraction of all-pairs shortest paths that pass through that node:
\begin{align}
c_B(v) = \sum_{s,t\in V} \frac{\sigma(s,t\, |\, v)}{\sigma(s,t)}
\label{bc}
\end{align}

\noindent where $V$ is the set of nodes, $\sigma(s,t)$ is the number of shortest paths, and $\sigma(s,t\, |\, v)$ is the number of those paths passing through some node $v$ other than $(s,t)$ \cite{b61}.

\CM{1,667}{487}{419}{1,751}{Confusion matrix for the \textit{baseline} model predicting $\mathcal{G}_{unknown}$.}{baselinecm}

\CM{1,901}{213}{451}{1,759}{Confusion matrix for the \textit{3-feature} model predicting $\mathcal{G}_{unknown}$.}{3featcm}

We also point out that the computational expense of determining these additional features is quite low compared to calculated $J(G_i)$, so they add a relatively minor cost for each graph.
With these two additional features, the new features matrix $\textbf{X}$ for each individual graph has gone from $\textbf{X} \in \mathbb{R}^{n \times 1}$ to $\textbf{X} \in \mathbb{R}^{n \times 3}$.
The CM for this experiment is in \tref{3featcm}.
Here we have that by adding the two additional features, the model was able to achieve an accuracy of 85\%, \textit{Precision} of 89.9\%, \textit{Recall} of 80.8\%, \textit{F1 Score} of 85\%, and an \textit{MCC} of 70\%.
Therefore, the GDL model was able to make better predictions on the same $\mathcal{G}_{known}$ dataset as all metrics are the same or better.
We will include all three features going forward.

\subsection{Experiment 3: Number of Epochs \& Number of Known Graphs}
\label{exp3r}

The previous experiments assumed a large number of epochs, but it is desirable to have insights into how many are actually required to effectively train the model to reduce computation costs.
Here we will observe the training behavior up to 1,000 epochs and decide a limit for future experiments.

Additionally, using 80\% of the data for training, generally as was done in the previous experiments, will not be desirable, especially when the cost of each $J(G_i)$ is quite high.
Therefore, we will also explore in this experiment different percentage values of $\mathcal{G}_{known}$, which represents fewer circuits that are sized using the expensive Eq.~(\ref{eq:cir-opt}).
The goal is to determine (for this particular dataset) a rule for the relative size of $\mathcal{G}_{known}$ to $\mathcal{G}_{all}$ that maintains the suitable accuracy while reducing the overall computational burden (both through model training time and time to generate $\mathcal{G}_{known}$).
We will expect the model's performance to degrade as the amount of training data decreases.

\begin{figure}[t]
\centering
\includegraphics[scale=0.95]{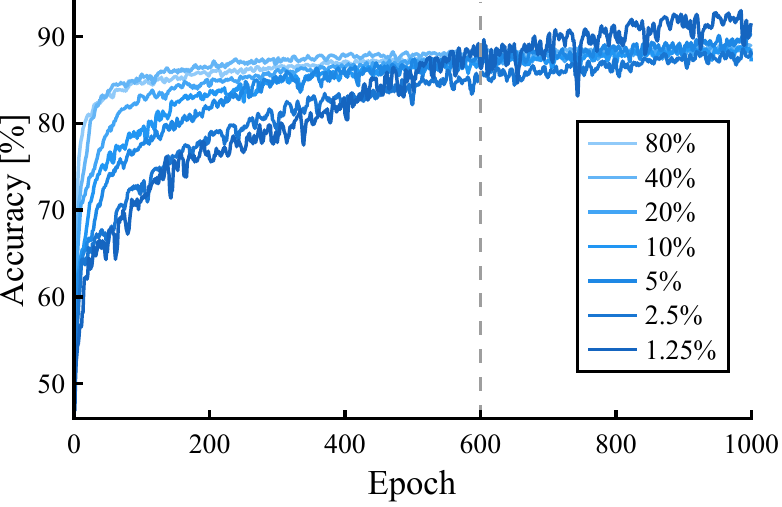}
\caption{Accuracy of seven different models during training using different percentages of the dataset to determine the approximate number of epochs stopping point of 600.}
\label{1000e}
\end{figure}

\begin{table*}[t]
\centering
\caption{The results for different values of $N_{known}$ with all metrics computed with respect to the unknown graphs $\mathcal{G}_{unknown}$.}
\label{exp3graphs}
\begingroup
\setlength{\tabcolsep}{6pt} 
\renewcommand{\arraystretch}{1.1} 
\begin{tabular}{rrrrrrrrrr}
\hline \hline
$N_{known}$ [\# of graphs] & 34,599 & 17,300 & 8,650 & 4,325 & 2,162 & 1,081 & 541 & 270 & 135 \\
\% of $N_{all}$ & $80\%$ & $40\%$ & $20\%$ & $10\%$ & $5\%$ & $2.5\%$ & $1.3\%$ & $0.6\%$ & $0.3\%$ \\
\hline
\textit{Training Time} [s] & 4,294 & 2,239 & 1,257 & 720 & 464 & 334 & 269 & 243 & 229 \\
\textit{Accuracy} [\%] & 83.80 & 83.99 & 82.70 & 81.73 & 78.68 & 76.15 & 72.28 & 66.70 & 60.69\\ 
\textit{Precision} [\%] & 84.10 & 88.37 & 80.29 & 79.11 & 77.18 & 76.51 & 70.97 & 70.54 & 67.70 \\
\textit{Recall} [\%] & 83.22 & 78.52 & 86.85 & 86.30 & 81.37 & 75.49 & 75.41 & 57.39 & 40.90\\
\textit{F1 Score} [\%] & 83.66 & 83.15 & 83.44 & 82.55 & 79.22 & 75.99 & 73.13 & 63.29 & 51.00\\
\textit{MCC} & 0.68 & 0.68 & 0.66 & 0.64 & 0.57 & 0.52 & 0.45 & 0.34 & 0.23\\
\hline \hline
\end{tabular}
\endgroup
\end{table*}

\subsubsection{Number of Epochs.}~Seven different models were training using different $N_{known}$ spaced between 1.25\% and 80\%.
Observing the results in \fref{1000e}, there is a clear distinction around 600 epochs where many of the accuracy metrics plateau.
For some of the smaller training sets, accuracy continues to increase beyond this point, but it was observed that this was mostly overfitting to the small training dataset.
Therefore, a 600 epoch limit is established and will generally result in models that are suitably trained without wasting additional iterations.

\subsubsection{Number of Known Graphs.}~
Using the same setup, we can explore trade-offs in the number of graphs (represented as a percentage here) included in $\mathcal{G}_{known}$.
Reducing $N_{known}$ has the consequence that $N_{unknown}$ increases, per Eq.~(\ref{eq:N}), so more graphs will need their classifications predicted.
Therefore, this experiment aims to see what minimum amount of data is required for the models to retain a high level of accuracy.

The results are shown in \tref{exp3graphs} with all metrics computed using the unknown graphs $\mathcal{G}_{unknown}$.
As we might expect, the larger $N_{known}$ is, the higher the model's accuracy. 
Now, simply reducing the amount of data in half from 80\% to 40\%, we still achieve a high accuracy of $\approx$83\% and similar \textit{MCC} of 0.68.
However, a key takeaway is the training time.
The 80\% model took 4,294 s (or about 72 min), while the 40\% model took only 37 minutes; the training time in half by cutting the data in half.
We should also consider the impact on computational cost required to construct $\mathcal{G}_{known}$; 40\% will nominally take half as long to determine all required $J(G_i)$.
Both of these factors are important in understanding what \% of $N_{all}$ is desirable when balancing total computational cost and model effectiveness (e.g., accuracy).
Looking into the smaller $N_{known}$ values, the results are also as expected: the smaller $N_{known}$, the lower many of the scores.
At the limit of the runs, only 135 graphs (0.3\%), the model was only able to achieve an accuracy of $\approx$60\% with a much lower \textit{MCC} of 0.23.

\begin{figure}[t]
\centering
\includegraphics[scale=0.95]{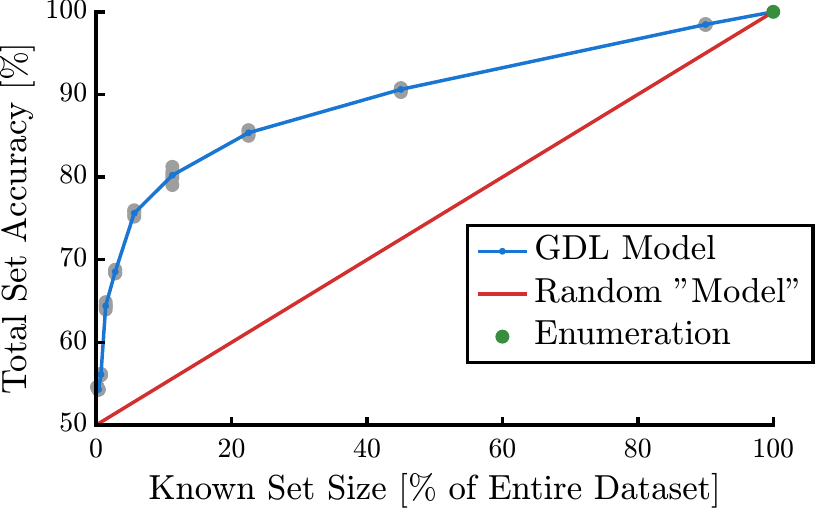}
\caption{Total set accuracy scores averaged over five runs for different known set sizes $N_{known}$.}
\label{fig:accvar}
\end{figure}

\begin{figure*}[t]
\centering
\includegraphics[scale=0.95]{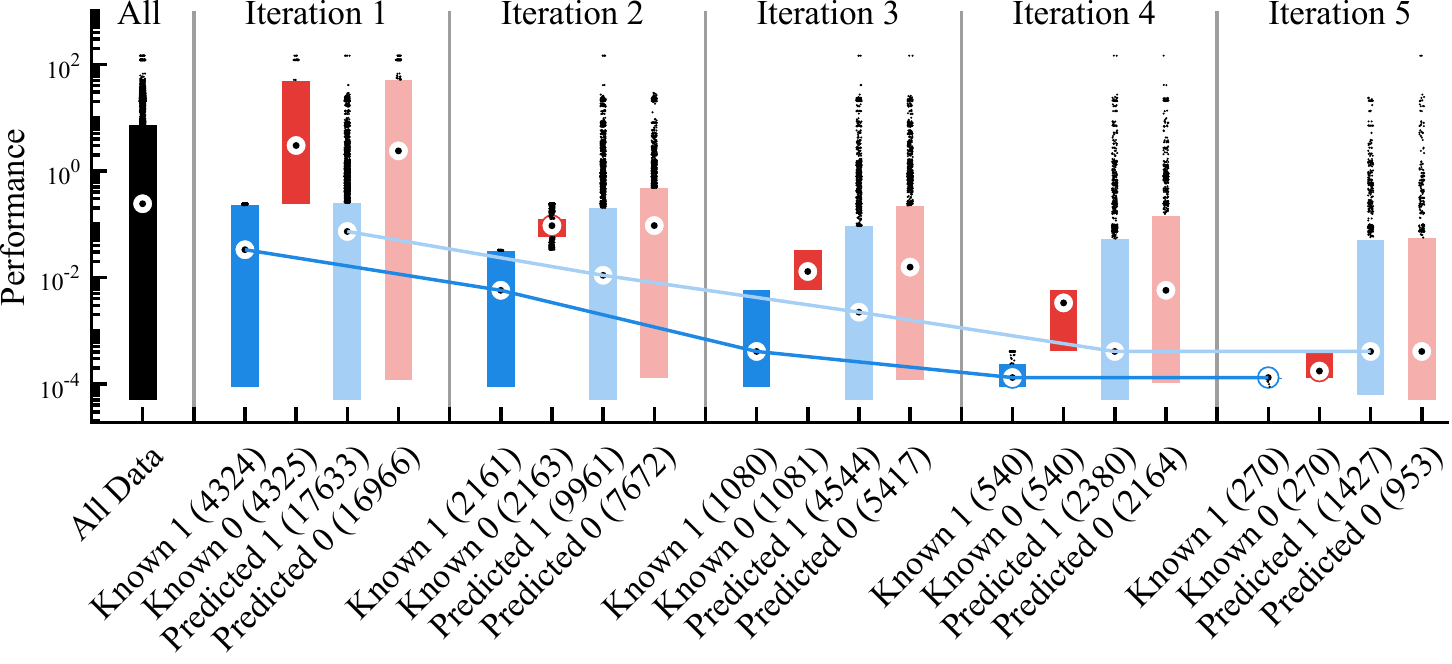}
\caption{Five iterations of the approach described in \sref{exp4r} for retraining a GDL model based on successive performance minimizing-focused subsets of the graphs (with the numbers indicating the number of graphs in the respective set) along with the distribution of all the data points $J(\mathcal{G}_{all})$ for comparison.}
\label{fig:iterations}
\end{figure*}

\begin{figure}[t]
\centering
\includegraphics[scale=0.95]{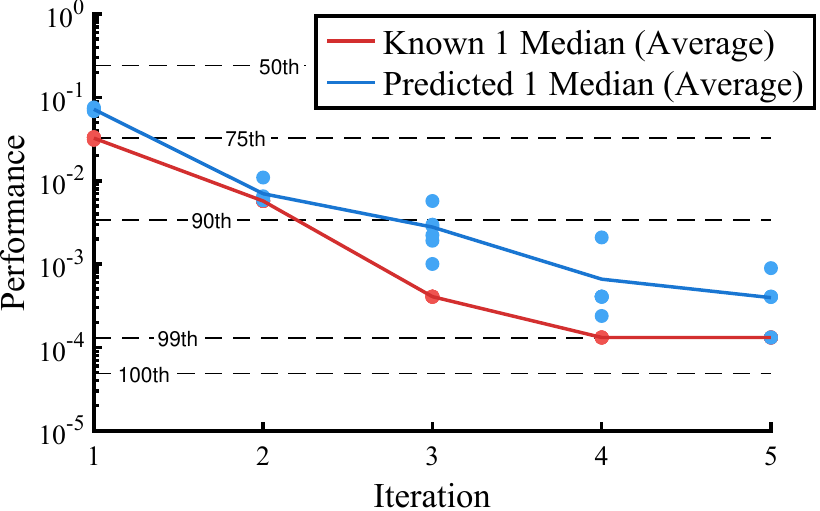}
\caption{Median values of the ``Known 1'' and ``Predicted 1'' sets averaged over six runs using the iterative GDL classification approach with different performance percentiles for $J(\mathcal{G}_{all})$ shown.} 
\label{fig:iterations-stats}
\end{figure}

To better understand the trade-offs in using GDL models for the case study, we visualize the results of the metric \textit{Total Set Accuracy} from Eq.~(\ref{eq:total-set-acc}) in \fref{fig:accvar}.
As there are several stochastic elements to the approach, the mean of five different runs is shown with the different gray dots indicating the values for the individual runs.
Overall, the variation is relatively smaller between runs.
In addition to this curve, a theoretical random ``model'' is added where we assume that all known values are predicted correctly, and all others are randomly assigned a good/bad classification.
Finally, when the known set size is 100\% of the dataset, then we have enumerated all graphs, so the total set \textit{Total Set Accuracy} is naturally 100\%.
Therefore, the GDL should be compared with respect to these additional standards.

Overall, we see the GDL model greatly outperforms the random model between around 1\%--40\%, but the gap begins to close as more graphs are known and all approach the 100\%/100\% enumeration point as $N_{known}$ increases.
Therefore, a general recommendation might be to select $N_{known}$ as 20\% of $N_{all}$, but this value certainly can be problem-specific and depend on the designer's preferences to balance \textit{Total Set Accuracy} vs. computational costs (which is generally proportional to $N_{known}$).

\subsection{Experiment 4: Iterative GDL Classification}
\label{exp4r}

Even if with perfect classification, the approach outlined so far would only result in determining the top 50\% performing graphs in $\mathcal{G}_{all}$.
While this outcome can undoubtedly be useful in practice, it is generally desirable to narrow down the graphs further to the best-performing.
However, there is no reason we only need to construct a single classification model.

In this experiment, we seek a smaller, better median performance set of graphs from $\mathcal{G}_{all}$ by iteratively constructing GDL models with these steps:
\begin{enumerate}[nolistsep]

\item Set $k=1$ and create an initial $\mathcal{G}^{k}_{known}$. 

\item Create a GDL model $m^k(G_i)$ using $\mathcal{G}^{k}_{known}$, which is naturally broken into sets ``Known 1'' and ``Known 0'' based on the median $J$ value of $\mathcal{G}^{k}_{known}$.

\item Predict the classes of the $\mathcal{G}^{k}_{unknown}$ using $m^k(G_i)$, creating ``Predicted 1'' and ``Predicted 0'', which are sets of graphs predicted to be good (1) or bad (0), respectively. 

\item The goal is to identify good graphs, so we set $\mathcal{G}^{k+1}_{known}$ equal to ``Known 1'' and $\mathcal{G}^{k+1}_{unknown}$ equal to ``Predicted 1'' (and the remaining graphs are removed under the assumption that they are bad).

\item Set $k \rightarrow k + 1$ and repeat Step 2 until $k = n$.
    
\end{enumerate}

\noindent At each iteration, the size of $\mathcal{G}^{k}_{known}$ will be halved in a way that decreases its median performance value, so the good/bad classification threshold of $m^k(G_i)$ should shift as well.

The results of this experiment initialized with 20\% of the data known are summarized in \fref{fig:iterations}.
First, the leftmost box plot shows the distribution of $J(\mathcal{G}_{all})$.
The next two box plots show the splitting of $\mathcal{G}_{known}$ based on the classification threshold at iteration 1.
The final two box plots in iteration 1 are the predicted 1s and 0s based on the GDL model; note that, while there are certainly incorrectly identified graphs, the medians are quite separated and consistent with their ``Known'' counterparts.
The subsequent box plots show the results for each iteration, up to $n=5$.
We observe at iteration 5, with $N_{known} = $ 540 graphs now, the classification is quite poor as the ``Predicted'' box plots are similarly distributed. 
The trend lines for the two medians also convey this point, with substantial decreases until the fifth iteration.
This behavior is perhaps expected because we are not adding any new data from the initial $\mathcal{G}^1_{known}$, and the number of graphs for training decreases.
Adding more data at each iteration based on ``Predicted 1'' is future work.

Assessing the outcome at the end of iteration 4 (since the fifth did not classify well), there are a total of 2,920 graphs in the final ``good'' set with a median of $3 \times 10^{-4}$, substantially lower than the initial classification median of $2 \times 10^{-1}$.
Here we assume a designer would evaluate $J(G_i)$ for all the graphs in ``Predicted 1''; thus, there would be a total of 11,029 graphs that were optimized. 
For this final set, \textit{all} of the top 10, 87 of the top 100, and 727 of the top 1000 graphs still remain at 25\% the computational cost of complete enumeration.
Furthermore, compared to randomly sampling 11,029 graphs from $\mathcal{G}_{all}$, the iterative GDL model results greatly exceed the expected means of 2.25, 22.51, and 225.13 for the top 10, 100, and 1000 graphs being included in the random set, respectively.

To better understand the statistical significance of this iterative approach, five more randomized runs were completed with different samplings of $\mathcal{G}^1_{known}$. 
Over these runs, the average graphs that would be known or ``optimized'' was 11282.2 graphs (with a 492.8 standard deviation).
For the top 100 graphs, 88.2 (2.5) remained compared to an expected value of 22.96.
Finally, for the top 1000 graphs, 751.2 (39.5) remained compared to an expected value of 229.6.
Furthermore, the median values of the ``Known 1'' and ``Predicted 1'' graph sets for each iteration are shown in Fig.~\ref{fig:iterations-stats} averaged over the six runs.
Here we still see a decrease in the median values indicating the iterative GDL approach is narrowing down the set of potentially ``good'' graphs to the higher percentiles of performance. In fact, at iteration 4, all medians for ``Predicted 1'' are better than the 90th percentile, with the mean closer to the 95th.
However, as previously discussed in Fig.~\ref{fig:iterations}, the benefits from the 5th iteration are minimal.

As indicated in Fig.~\ref{fig:accvar}, there are expected trade-offs when attempting to reduce computational costs by decreasing the required evaluations of $J(G_i)$ and GDL tasks.
Therefore, an additional study was conducted only reducing the starting $\mathcal{G}^{1}_{known}$. 
Initialized with only $10\%$ of $\mathcal{G}_{all}$, the iterative approach now concluded with 8118.2 (969.1) graphs on average needing to be optimized per Eq.~(\ref{eq:cir-opt}). 
Summarizing the results of five random runs, 9.2 (0.4) of the top 10 graphs remained, 85.2 (6.3) of the top 100 graphs remained, and 715.2 (76.9) of the 1000 graphs remained.
Another difference between the different starting $\mathcal{G}_{known}$ sizes (i.e., $20\%$ vs.~$10\%$) is total iterative GDL classification cost, including training the various GDL models at each iteration. 
The difference here is 45 vs. 36 minutes between 20\% vs.~10\% $\mathcal{G}^{1}_{known}$, respectively.
Since both summarized top graph outcomes are comparatively similar, additional investigations into using fewer graphs for $\mathcal{G}^{1}_{known}$ are critical to understanding trade-offs related to the total computational costs of the design study to meet different designer goals.

\section{CONCLUSION}
\label{conclusion}
This paper presents a Geometric Deep Learning (GDL) approach for classifying and down-selecting graph-based analog circuits towards sets of better-performing solutions based on their performance evaluation in an engineering design problem.
In the presented graph-design case study where all the graphs were known but not their expensive optimization-based performance values, we have demonstrated the potential capabilities of GDL in helping classify ``good'' and ``bad'' graphs based on limited performance data, as well as some insights into the key hyperparameters that need to be tuned.
Additionally, including the graph-based features eigenvalue centrality and betweenness centrality improved many of the key metrics.

The results showed that GDL can be used as an effective and efficient method for narrowing the graph sets for this case study, as the model achieved a total set classification accuracy of 80\% 
Furthermore, the iterative GDL classification approach identified 9.0 of the top 10,  88.2 of the top 100 graphs, and 751.2 out of the top 1000 at 25\% the computational cost of complete enumeration.

Key future work items include investigating iteratively adding new graphs to the dataset as mentioned in \sref{exp4r}, alternative approaches to predict the classes of $\mathcal{G}_{unknown}$ using the trained GDL model (changing what graphs would be removed at each iteration), and transfer learning on similar problems~\cite{b5}.
There is also the task of investigating this approach on other datasets in different engineering fields, such as ones with directed graphs and multiple performance metrics.
As is the case with many GNN-based approaches, there is still work to be done to explore the best neural network architecture and hyperparameters to reduce the data input requirements and improve model accuracy.
Furthermore, similar to Sec.~\ref{exp2r}, adding other graph-based features could enable more favorable outcomes (e.g.,~current-flow closeness centrality, closeness vitality, and harmonic centrality \cite{Koschutzki2005, boldi2013axioms}).
The promising results for this large dataset of small engineering graphs indicate a bright future for iterative classification-based GDL for a broader application in other engineering graph-design problems.


\renewcommand{\refname}{REFERENCES}
\bibliographystyle{config/asmems4}
\begin{mySmall}
\bibliography{References}
\end{mySmall}


\appendix
\section{TOOLS AND COMPUTING ARCHITECTURE}
\label{appa}
The primary tool used is PyTorch-Geometric (PyG) \cite{b16} because of the extensive applications that PyG can perform on graph data.
It is built on top of PyTorch \cite{b15}, an open-source machine learning framework, which also contains an extensive library of tools for model manipulation and data analysis.
Since both tools are used primarily with Python \cite{b57}, that will be the language of choice for the model.
We also include Networkx, a Python package for the creation and manipulation of complex networks \cite{b17}.
This decision is because PyG requires the data to be of a certain instance, whether a SciPy sparse matrix or a Trimesh instance; it needs to be in a form readable by PyG.
Networkx was chosen because of the information that can be added to the graph, which can be easily transformed into a PyG instance.
We then use Pandas \cite{b77} for its data organization capabilities to import the data and prepare it to be passed on to Networkx and, finally, PyG.
For a full list of the tools used and their versions, please see \tref{tooltab}.

All the tools were utilized on a personal workstation consisting of an Intel Core i9-9900k CPU @ 3.60GHz, 32GB installed memory, and an Nvidia GeForce RTX 2060 Super GPU.

\begin{table}[h]
\centering
\caption{A list of the primary tools and their versions.}
\label{tooltab} 
\begingroup
\setlength{\tabcolsep}{6pt} 
\renewcommand{\arraystretch}{1.1} 
\begin{tabular}{rl}
\hline \hline
\textbf{Tool} & \textbf{Version} \\
\hline
Python & 3.9\\
Networkx & 2.8.7\\
PyTorch & 1.12.1\\
PyTorch-Geometric & 2.1.0\\
SciPy & 1.9.1\\
Pandas & 1.5.0\\
\hline \hline
\end{tabular}
\endgroup
\end{table}

\end{document}